\documentclass[a4paper,fleqn]{cas-sc}
\usepackage{hyperref}
\usepackage[nameinlink]{cleveref}
\usepackage{booktabs}  
\usepackage[authoryear]{natbib}
\usepackage{graphicx} 
\usepackage{float}
\usepackage{algorithm}  
\usepackage{algpseudocode}
\usepackage{color}
\usepackage{utfsym}
\usepackage{setspace}

\usepackage{lineno}
\usepackage[utf8]{inputenc}
\usepackage{tcolorbox}
\tcbuselibrary{listingsutf8}
\usepackage{listings}
\usepackage{xcolor}

\definecolor{codebg}{rgb}{0.98,0.98,0.98}
\definecolor{mygray}{rgb}{0.5,0.5,0.5}
\definecolor{myblue}{rgb}{0.2,0.3,0.7}
\definecolor{mypurple}{rgb}{0.6,0.2,0.6}
\tcbuselibrary{listings, breakable}
\newtcblisting{mycode}{
    colback=codebg,
    breakable, 
    colframe=black!50,
    listing only,
    listing options={
        language=Python,
        basicstyle=\ttfamily\small,
        keywordstyle=\color{myblue},
        stringstyle=\color{mypurple},
        commentstyle=\color{mygray}\itshape,
        numbers=left,
        numberstyle=\tiny\color{gray},
        stepnumber=1,
        numbersep=6pt,
        showstringspaces=false,
        breaklines=true,
        tabsize=4,
        columns=fullflexible,
        escapeinside={(*@}{@*)}
    },
    sharp corners,
    boxsep=1mm,
    left=6mm,
    right=1mm,
    top=1mm,
    bottom=1mm,
    fonttitle=\bfseries,
}

\crefname{algorithm}{Algorithm~}{Algorithms~}
\Crefname{algorithm}{Algorithm~}{Algorithms~}
\crefname{table}{Table~}{Tables~}
\Crefname{table}{Table~}{Tables~}
\crefname{figure}{Fig.}{Figs.}
\Crefname{figure}{Fig.}{Figs.}
\crefname{equation}{Eq.}{Eqs.}
\Crefname{equation}{Eq.}{Eqs.}

\def\tsc#1{\csdef{#1}{\textsc{\lowercase{#1}}\xspace}}
\tsc{WGM}
\tsc{QE}
\tsc{EP}
\tsc{PMS}
\tsc{BEC}
\tsc{DE}

\usepackage{lineno}

\begin{document}
\let\WriteBookmarks\relax
\def\floatpagepagefraction{1}
\def\textpagefraction{.001}
\shorttitle{ }
\shortauthors{ }

\title [mode = title]{Fast ground penetrating radar dual-parameter full waveform inversion method accelerated by hybrid compilation of CUDA kernel function and PyTorch }

\author{Lei Liu}[type=author,
                        auid=000,bioid=1,orcid=0000-0002-1313-8186]
\credit{Coding, Writing}

\author{Chao Song}
\credit{Methodology, Funding acquisition, Supervision, Validation, Writing - review \& editing}

\author{Liangsheng He}
\credit{Coding, Validation}

\author{Silin Wang}
\credit{Coding}

\author{Xuan Feng}
\credit{Methodology, Review}

\author{Cai Liu}
\credit{Methodology, Review}

\address{State Key Laboratory of Deep Earth Exploration and Imaging, College of GeoExploration Science and Technology, Jilin University, Changchun, 130026, China}

\begin{abstract}
This study proposes a high-performance dual-parameter full waveform inversion framework (FWI) for ground-penetrating radar (GPR), accelerated through the hybrid compilation of CUDA kernel functions and PyTorch. The method leverages the computational efficiency of GPU programming while preserving the flexibility and usability of Python-based deep learning frameworks. By integrating customized CUDA kernels into PyTorch's automatic differentiation mechanism, the framework enables accurate and efficient inversion of both dielectric permittivity and electrical conductivity. Experimental evaluations on synthetic data and real wavefield data demonstrate that the proposed method achieves dual-parameter FWI for GPR data while maintaining high accuracy. Moreover, the framework is flexible and extensible, supporting optional regularization strategies such as total variation and multi-scale inversion. These features make the proposed approach a practical and scalable framework for rapid GPR-based subsurface imaging in applications including civil engineering, environmental monitoring, and geophysical exploration.
\end{abstract}


\maketitle 

\printcredits

\doublespacing

\section{Introduction}
\label{intro}

Ground penetrating radar (GPR) is a non-destructive geophysical technique that employs electromagnetic waves across from 10 MHz to 4 GHz range to image subsurface structures by analyzing reflections caused by permittivity \citep{daniels2004ground}. Some studies have demonstrated the utility of GPR in landfill monitoring applications, including the estimation of in situ water content \citep{yochim2013estimating}, identification of preferential leachate migration pathways via fractured media \citep{pujari2007assessment}, and delineation of contamination plumes beyond landfill boundaries \citep{porsani2004use}. 

Full waveform inversion (FWI) significantly enhances imaging resolution by reconstructing both dielectric permittivity and electrical conductivity through comprehensive waveform analysis \citep{lavoue2014two,song2021wavefield}. However, the high computational demands of FWI present substantial challenges. Conventional FWI frameworks, which involve solving Maxwell's equations repeatedly \citep{ernst2007full}, require intensive computational resources, especially when dealing with high-frequency data (e.g., >1 GHz), which necessitates fine spatial discretization \citep{giannopoulos2005modelling}. Moreover, simultaneous inversion of multiple parameters (e.g., relative permittivity and conductivity) will introduce extra computational cost \citep{hernandez2023use}. The integration of advanced techniques such as machine learning-assisted regularization \citep{wang2020gpr} and multi-physics fusion \citep{feng2017quantitative} has further intensified the computational burden of FWI. These computational challenges hinder the development of FWI for GPR data, such as civil infrastructure assessment, environmental monitoring, and rapid subsurface anomaly detection \citep{lopera2007filtering}. 

Recent progress in Graphics Processing Unit (GPU) parallelization, has shown considerable promise in accelerating FWI by several orders of magnitude. Compute Unified Device Architecture (CUDA)-based acceleration techniques have been widely adopted in geophysical forward modeling and inversion tasks. In seismic exploration, various studies have successfully implemented numerical methods, such as the finite difference time domain (FDTD) \citep{weiss2013solving} and finite element method (FEM) \citep{komatitsch2010high} on GPU platforms to improve the efficiency of wavefield simulation and gradient computation \citep{jia2016high,rietmann2012forward}. Deepwave, a flexible seismic modeling and inversion package proposed by \citet{richardson_alan_2023}, embeds wave propagators as PyTorch modules capable of running on both CPUs and GPUs, with significant efficiency improvements achieved through CUDA kernel function acceleration. Similarly, \citet{han2016simulation} decomposed frequency-sweeping tasks in magnetotelluric (MT) forward modeling into independent CUDA kernels, attaining speedups of 18.5 to 21.7 times. In electromagnetic exploration, especially for GPR simulations, several studies have converted 2D and 3D electromagnetic wave propagation schemes into CUDA-parallelized kernels, substantially reducing simulation time. For example, \citet{feng2018fast} achieved a 10× speedup in 2D GPR dual-parameter forward modeling using GPU-optimized conjugate gradient methods, while \citet{wang20193d} reduced computational costs of FWI through variable-grid strategies. Notably, GprMax integrates pyCUDA and CUDA kernel functions to accelerate the forward modeling process \citep{warren2019cuda}. Additionally, \citet{wang2025gpr} introduced GPR-FWI Py, a 2D GPR FWI software package implemented in pure Python, optimized for high-performance CPU-based computing. However, there is still a significant gap in computational efficiency compared to GPU-based methods. These developments highlight the critical need of adopting GPU for accelerating FWI, thereby facilitating its wavefield applications for real-time subsurface diagnostics.

Despite these advances, few studies have effectively combined CUDA kernel functions with high-level Python interfaces to support efficient FWI for GPR data. This study proposes a fast GPR FWI framework based on hybrid compilation of CUDA kernels and Python, which substantially improves the efficiency of both forward and adjoint modeling. Furthermore, the proposed method offers robust support for dual-parameter FWI using automaitc differentiation.

Synthetic and real data tests demonstrate that the proposed method achieves outstanding computational efficiency, ease of use, and high inversion quality. 

On a computer equipped with an RTX 4090 graphics card, in a two-parameter inversion scenario with a grid size of about 200×100, the total running time for two hundred epochs is less than five minutes, and each training cycle is completed in less than two seconds. To further enhance computational efficiency, several optimization strategies are incorporated to reduce the number of training epochs without sacrificing model accuracy. After the initial compilation, the inversion process can be flexibly configured through input parameters, with key functionalities controlled via Boolean flags. This streamlined interface facilitates efficient experimentation with various inversion settings. The method achieves satisfactory inversion results with minimal hyperparameter tuning—such as adjusting the learning rate or choosing suitable loss functions—while increasing the number of training batches contributes to improved model fidelity. 

These characteristics demonstrate that the proposed approach is not only computationally efficient but also extensible and well-suited for practical applications in GPR FWI.

\section{Methodology}

\subsection{Full waveform inversion (FWI)}

The time-domain electromagnetic wavefield canbe simulated by solving the equationscorresponding to Faraday's Law and Ampère-Maxwell Law \citep{taflove2005computational}.:

\begin{align}
\nabla \times \mathbf{E} &= -\mu \frac{\partial \mathbf{H}}{\partial t}, \label{eq:faraday}
\end{align}
\begin{align}
\nabla \times \mathbf{H} &= \sigma \mathbf{E} + \varepsilon \frac{\partial \mathbf{E}}{\partial t}, \label{eq:ampere}
\end{align}
where $\mathbf{\nabla}$ denotes vector differential operator,
$\mathbf{E}$ represents electric wavefield strength (V/m),
$\mathbf{H}$ represents magnetic wavefield strength (A/m),
$\mathbf{\mu}$ is magnetic permeability (H/m, usually free space magnetic permeability),
$\mathbf{\varepsilon}$ is permittivity (F/m),
$\mathbf{\sigma}$ is conductivity (S/m).

In a medium with non-zero conductivity, the electric wavefield will attenuate due to current loss. Thus, in FDTD simulation, the conductivity term needs to be included $\sigma \mathbf{E}$ on the right-hand side of Ampère's law. This describes Ohmic losses, which are particularly critical in GPR simulations as the subsurface media are typically lossy dielectric (such as wet soil, clay, etc.).

The equations for discretizing the above equations for two-dimensional electromagnetic wavefield of Transverse Magnetic (TM) mode are given by:
\begin{align}
E_z^{n+1}(i,j) =\; & \left( 1 - \frac{\sigma(i,j) \Delta t}{\varepsilon(i,j)} \right) E_z^{n}(i,j) \nonumber \\
                  +\; & \frac{\Delta t}{\varepsilon(i,j)} \left[
\frac{H_y^{n+\frac{1}{2}}(i,j) - H_y^{n+\frac{1}{2}}(i-1,j)}{\Delta x}
- \frac{H_x^{n+\frac{1}{2}}(i,j) - H_x^{n+\frac{1}{2}}(i,j-1)}{\Delta y}
\right], \label{eq:ez_explicit}
\end{align}
\begin{align}
H_x^{n+\frac{1}{2}}(i,j) &= H_x^{n-\frac{1}{2}}(i,j) 
- \frac{\Delta t}{\mu(i,j)} \cdot \frac{E_z^n(i,j+1) - E_z^n(i,j)}{\Delta y}, \label{eq:hx_update} \\
H_y^{n+\frac{1}{2}}(i,j) &= H_y^{n-\frac{1}{2}}(i,j) 
+ \frac{\Delta t}{\mu(i,j)} \cdot \frac{E_z^n(i+1,j) - E_z^n(i,j)}{\Delta x}. \label{eq:hy_update}
\end{align}

The objective function of GPR FWI is expressed in the form of least squares error:
\begin{equation}
\mathcal{J}(\varepsilon, \sigma) = \frac{1}{2} \sum_{r,t} \left[ E^{\text{cal}}(\mathbf{x}_r, t;\mathbf{\varepsilon},\mathbf{\sigma}) - E^{\text{obs}}(\mathbf{x}_r, t) \right]^2, \label{eq:objective}
\end{equation}
where 
$\mathbf{E^{\text{cal}}}$ and $\mathbf{E^{\text{obs}}}$ represent simulated and observed electric wavefield data,
$\mathbf{\mathbf{x}_r}$ denotes receiver location,
$\mathbf{t}$ is the time dimension. 

We introduce Lagrange multipliers $\mathbf{\lambda}_E$ and $\mathbf{\lambda}_H$ construct the following Lagrange function:
\begin{equation}
\mathcal{L} = \mathcal{J} 
+ \int_0^T \int_\Omega \boldsymbol{\lambda}_H \cdot \left( \mu \frac{\partial \mathbf{H}}{\partial t} + \nabla \times \mathbf{E} \right) \, d\mathbf{x}dt
+ \int_0^T \int_\Omega \boldsymbol{\lambda}_E \cdot \left( \varepsilon \frac{\partial \mathbf{E}}{\partial t} + \sigma \mathbf{E} - \nabla \times \mathbf{H} - \mathbf{S} \right) \, d\mathbf{x}dt,
\label{eq:lagrangian}
\end{equation}
where $\mathcal{J}$ is the data-fitting objective,
$\mathbf{S}$ is the excitation source.

Make a variation on ${E}$ and ${H}$ and let $\delta \mathcal{J}=0$ , then we can derive the adjoint wavefield equation for back propagation.
\begin{align}
\mu \frac{\partial \boldsymbol{\lambda}_H}{\partial t} &= + \nabla \times \boldsymbol{\lambda}_E \label{eq:adj_H} ,
\end{align}
\begin{align}
\varepsilon \frac{\partial \boldsymbol{\lambda}_E}{\partial t} + \sigma \boldsymbol{\lambda}_E &= - \nabla \times \boldsymbol{\lambda}_H + \frac{\partial \mathcal{J}}{\partial \mathbf{E}} \label{eq:adj_E},
\end{align}

Termination time condition:
\begin{align}
\boldsymbol{\lambda}_E(T) = \boldsymbol{\lambda}_H(T) = 0, \label{eq:time0}
\end{align}

where $\boldsymbol{\lambda}_E(T)$ and $\boldsymbol{\lambda}_H(T)$ is the adjoint variable of electric and magnetic fields. $T$ is the end time of the entire inversion time window, which is the final time of the wave field propagation simulation.

Stability conditions (Courant-Friedrichs-Lewy conditions, CFL conditions) need to be met during forward modeling:
\begin{equation}
\Delta t \leq \frac{1}{c \sqrt{ \left( \frac{1}{\Delta x^2} + \frac{1}{\Delta y^2} \right) }}, \label{eq:cfl}
\end{equation}
In \cref{eq:cfl}, $\Delta t$ is the time step, $\Delta x$ is the spatial step in the x direction, $\Delta y$ is the spatial step in the y direction, and $c$ is the speed of light.

We take partial derivatives of the model parameters. Based on the chain rule and $\delta\mathcal{L}/\delta\varepsilon=0$. We can obtain the gradient of conductivity:
\begin{equation}
\frac{\partial \mathcal{J}}{\partial \sigma(\mathbf{x})} = \int_0^T \boldsymbol{\lambda}_E(\mathbf{x}, t) \cdot \mathbf{E}(\mathbf{x}, t) \ dt ,\label{eq:grad_sigma}
\end{equation}
gradient of relative permittivity:
\begin{equation}
\frac{\partial \mathcal{J}}{\partial \varepsilon(\mathbf{x})} = \int_0^T \boldsymbol{\lambda}_E(\mathbf{x}, t) \cdot \frac{\partial \mathbf{E}(\mathbf{x}, t)}{\partial t} \, dt. \label{eq:grad_eps}
\end{equation}

We use the Adam optimizer for updates, which combines the ideas of Momentum and RMSProp. The updates of relative permittivity and conductivity are as follows:
\begin{equation}
\varepsilon_r^{(t+1)} = \varepsilon_r^{(t)} - \alpha_{\varepsilon} \cdot \frac{\hat{m}_t^{(\varepsilon)}}{\sqrt{\hat{v}_t^{(\varepsilon)}} + \varepsilon},
\end{equation}
\begin{equation}
\sigma^{(t+1)} = \sigma^{(t)} - \alpha_{\sigma} \cdot \frac{\hat{m}_t^{(\sigma)}}{\sqrt{\hat{v}_t^{(\sigma)}} + \varepsilon}.
\end{equation}
where $\alpha_{\varepsilon}$ and $\alpha_{\sigma}$ denote the learning rates for the relative permittivity $\varepsilon_r$ and electrical conductivity $\sigma$, respectively. The terms $\hat{m}_t^{(\cdot)}$ and $\hat{v}_t^{(\cdot)}$ represent the bias-corrected first- and second-order moment estimates of the gradient, while $\varepsilon$ is a small positive constant introduced to ensure numerical stability by avoiding division by zero.

\begin{figure}
\centering
\includegraphics[width=0.75\textwidth]{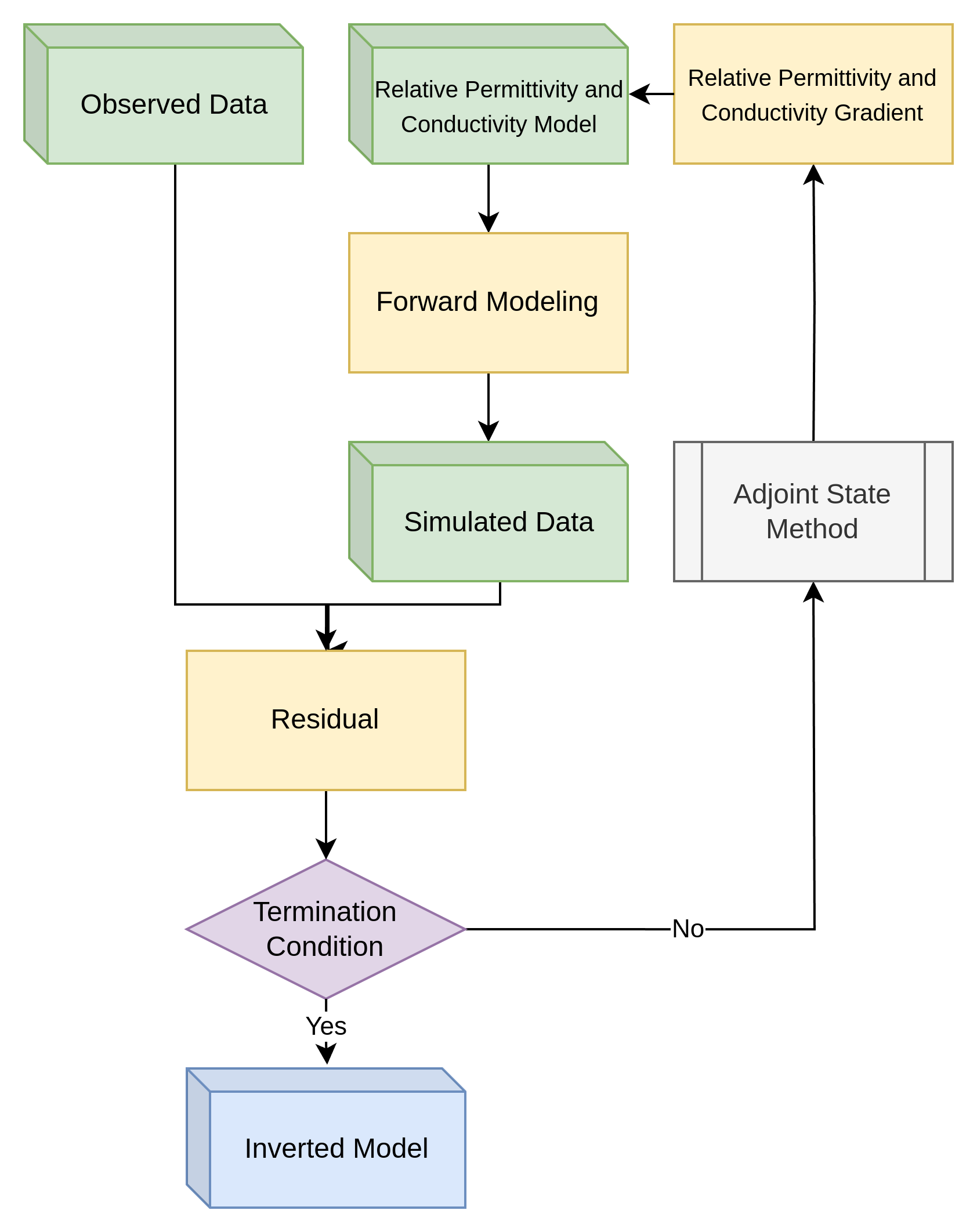}
\caption{Workflow of GPR FWI.}
\label{fig:fwi}
\end{figure}

\cref{fig:fwi} shows the basic process of GPR FWI. The results by forward modeling the initial model and observation data obtained are used to obtain the update gradient using the adjoint state method to guide the subsequent model update\citep{plessix2006review,song2023weighted}. The forward modeling results of the updated model gradually approach the observed data until the termination condition is reached.

\subsection{Compute Unified Device Architecture (CUDA) kernel}

A CUDA kernel is a function defined within the CUDA parallel computing framework proposed by NVIDIA. It is executed on the GPU and is capable of launching thousands of threads to perform computations concurrently.

When a kernel is invoked from the host (CPU) side, the syntax is typically as follows:
\begin{mycode}
myKernel<<<gridDim, blockDim>>>(...);
\end{mycode}

This call launches \texttt{gridDim} $\times$ \texttt{blockDim} threads on the GPU to execute the function myKernel() in parallel. Each thread is assigned a unique global identifier computed as:

\begin{equation}
globalThreadId = blockIdx.x \times blockDim.x + threadIdx.x \label{eq:threadid}
\end{equation}
In \cref{eq:threadid}, \texttt{threadIdx.x} denotes the thread's index within a block, \texttt{blockIdx.x} represents the index of the block within the grid, and \texttt{blockDim.x} specifies the number of threads per block. Using this global identifier, each thread can process a unique subset of the data, enabling fine-grained parallelism.

\begin{figure}
\centering
\includegraphics[width=0.75\textwidth]{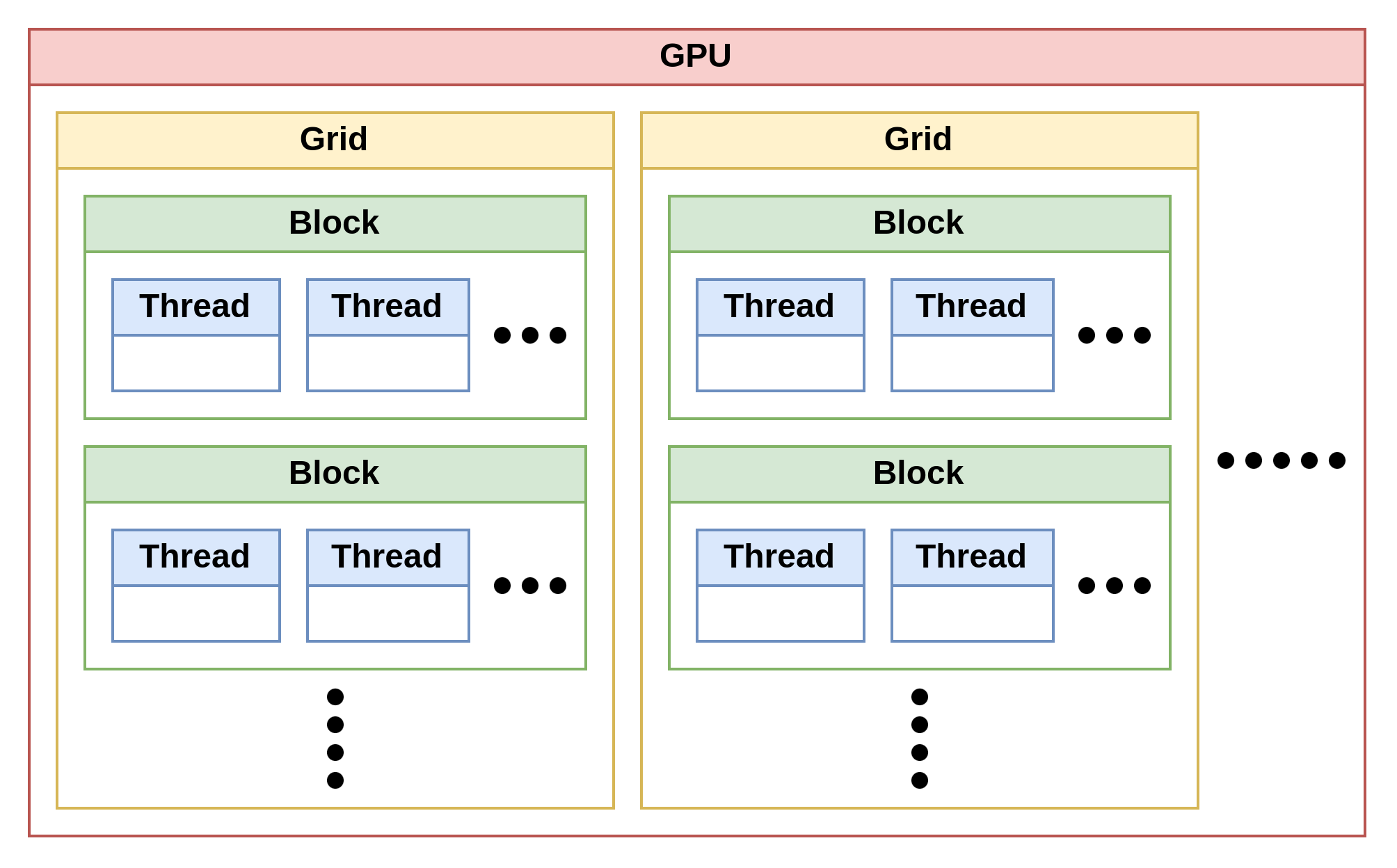}
\caption{Kernel parallel call structure.}
\label{fig:thread}
\end{figure}

The parallel execution hierarchy of CUDA kernels is illustrated in \cref{fig:thread}. In this structure, the GPU consists of multiple grids, each grid comprises several blocks, and each block contains multiple threads. Although only a subset is shown in the \cref{fig:thread}, this hierarchical organization supports scalable parallel execution across large datasets.

Importantly, GPU threads are not executed sequentially; instead, they are grouped into batches known as warps, each consisting of 32 threads. These warps are scheduled and executed in parallel under the Single Instruction, Multiple Thread (SIMT) model, which allows for efficient utilization of GPU hardware resources.

\subsection{Mixed compilation}

PyTorch is an open-source deep learning framework proposed by Facebook's AI research team \citep{paszke2017automatic}. It is widely adopted in both academia and industry due to its dynamic computational graph execution, flexible tensor operations, GPU acceleration, and built-in automatic differentiation (Autograd) \citep{paszke2019pytorch}. While using PyTorch's Autograd for FWI simplifies gradient computation, it also presents notable drawbacks—particularly in terms of computational efficiency and memory overhead, which can hinder its applicability to large-scale problems.

FWI for GPR is computationally intensive by nature, requiring repeated numerical solutions of Maxwell's equations and iterative parameter optimization. To address the efficiency bottlenecks inherent in FWI, integrating CUDA kernel functions with PyTorch offers a promising and efficient solution. This hybrid strategy leverages the fine-grained, low-level parallelism of CUDA alongside the high-level programmability and flexibility of PyTorch, significantly accelerating electromagnetic wavefield simulation. By embedding custom CUDA kernels within the forward and backward methods of $torch.autograd.Function$, the framework ensures compatibility with PyTorch's automatic differentiation engine, allowing direct and efficient optimization of physical parameters, such as relative permittivity and electrical conductivity. This integration not only enables fine-grained control over memory access patterns and thread scheduling but also supports the seamless adoption of PyTorch's native loss functions and optimization algorithms. As a result, the approach balances computational efficiency and usability, facilitating rapid and scalable GPR FWI implementations suitable for real-world subsurface imaging tasks. The interface between PyTorch and the CUDA modules is implemented via  \texttt{ctypes}, ensuring efficient and flexible communication between the two environments.

\section{Algorithm and implementation}
\subsection{Algorithm Flow}

\begin{figure}
\centering
\includegraphics[width=0.9\textwidth]{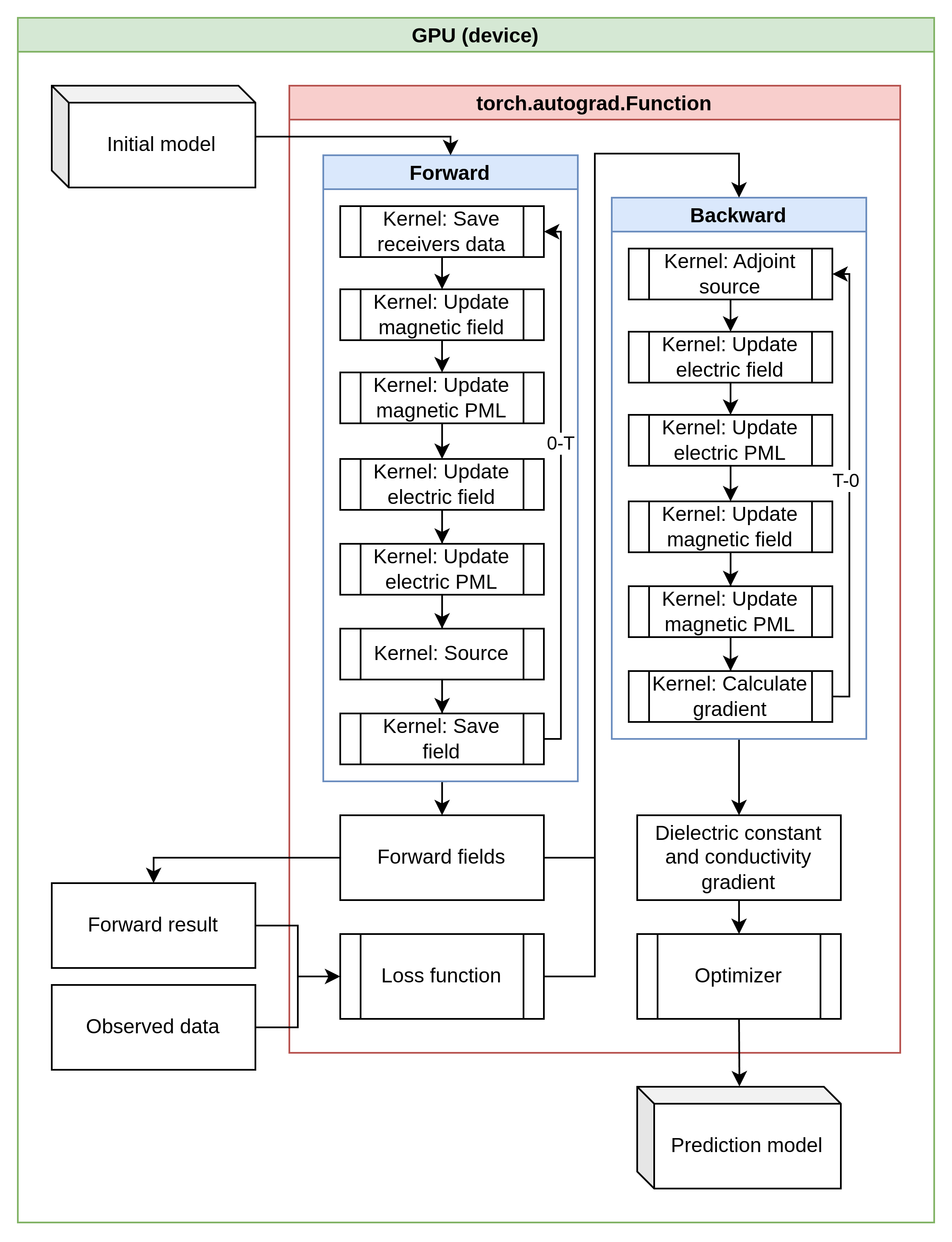}
\caption{Workflow of proposed method.}
\label{fig:deepgpr}
\end{figure}

The complete workflow is illustrated in \cref{fig:deepgpr}. All computational data are stored in GPU memory throughout the process. Upon completion of the forward modeling, the simulated results are output; after the inversion process, the final inverted model is generated.

For the first-time use, CUDA toolkit compatible with their hardware and the corresponding version of PyTorch with CUDA support should be installed. Once the required Python dependencies are installed, the CUDA kernel functions must be compiled. A provided Makefile automates the compilation process. These steps are only necessary once, unless the hardware is changed or the kernel code is modified.

All configurable parameters for our method are listed in Table 1, including their names, data types, descriptions, and whether each parameter is mandatory or optional.

\begin{figure}
\centering
\includegraphics[width=1\textwidth]{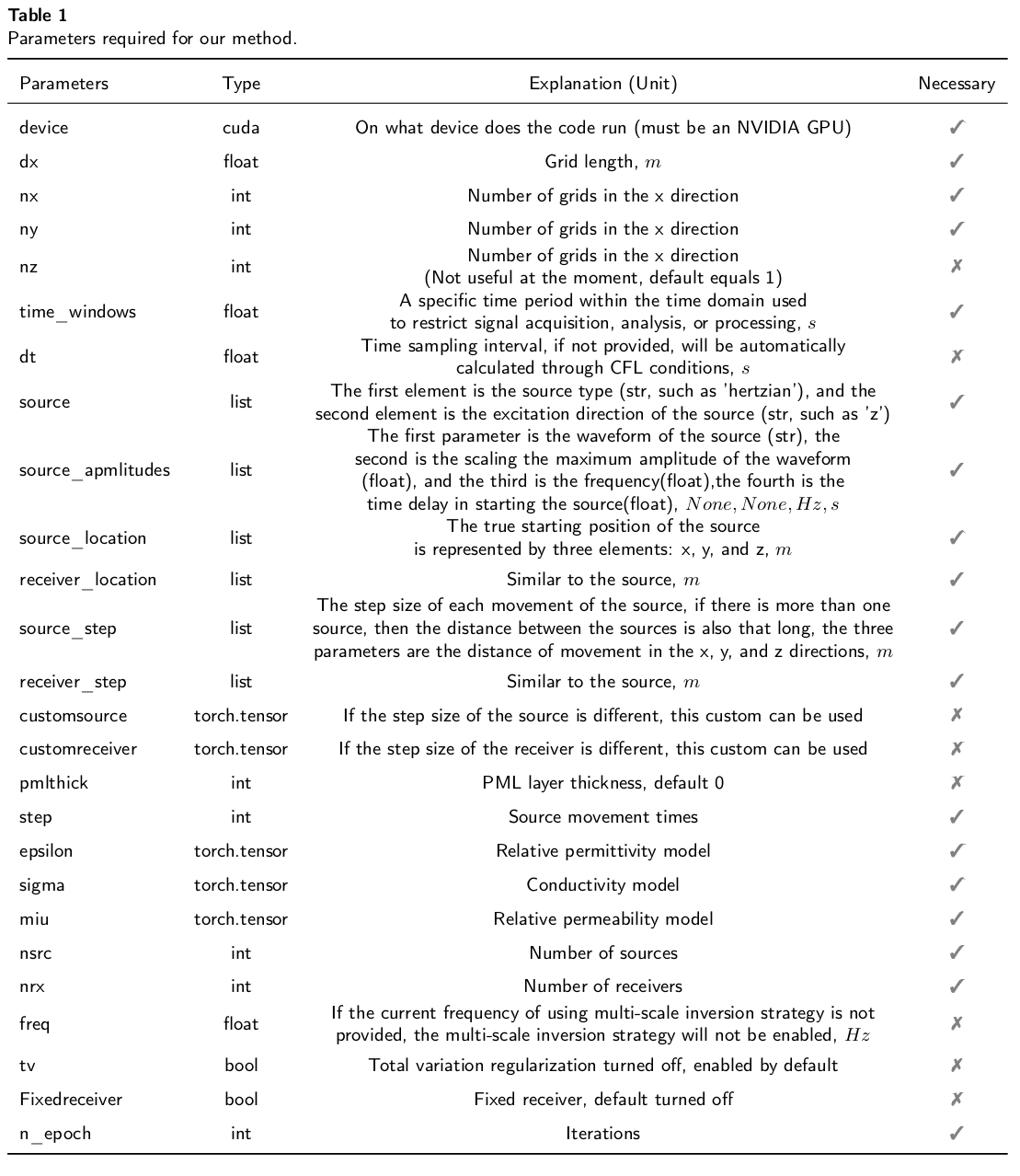}
\end{figure}

Once all parameters are configured, the FWI can be performed using the provided script. The inversion process closely resembles the training of deep learning neural networks, and the included script serves as a simplified demonstration of key functionalities.
\begin{mycode}
...
# Add requires_grad attribute to the parameters that need to be inverted
epsilon.requires_grad_()
sigma.requires_grad_()
# Set up optimizer to allocate different learning rates, taking Adam optimizer as an example
optimiser = torch.optim.Adam([
    {'params': epsilon, 'lr': 1},
    {'params': sigma, 'lr': 0.0001}
])
# Define the loss function, here we take MSE as an example
loss_fnmse = torch.nn.MSELoss()
# Save start time
start_time = time.time()
# Loop through all training epochs
for epoch in range(n_epochs):
    # Reset gradients before each backward pass
    optimiser.zero_grad()
    # Run the forward model to simulate wave propagation and get inverted data
    result, tt = compute(device, dx, ny, nx, nz, time_windows, source, source_location, 
                           source_step, receiver_location, receiver_step, pmlthick, 
                           source_amplitudes, step, epsilon, sigma, miu, nsrc, nrx, freq, dt)
    # Compute MSE loss between prediction and observation
    loss = loss_fnmse(result, observed_data) 
    # Perform backpropagation to compute gradients
    loss.backward()
    # Zero the gradients of model parameters near the source location
    if epsilon.requires_grad:
        epsilon.grad[:, :pmlthick+1, :] = 0
    if sigma.requires_grad:
        sigma.grad[:, :pmlthick+1, :] = 0
    # Apply gradient descent to update epsilon and sigma
    optimiser.step()
    # Compute time used for this epoch
    epoch_time = time.time() - epoch_start 
    # Compute total time since training started 
    total_time = time.time() - start_time  
    print(f"Epoch {epoch}, Loss: {loss:.6f}, Freq: {freq/1000000:.0f}MHz, Epoch Time: {epoch_time:.2f}s, Total Time: {total_time:.2f}s")
    # Set the frequency for saving intermediate results
    save_epoch=30
    # Extract the valid region (excluding PML) and save epsilon and sigma
    if epoch % save_epoch == 0:
        epsilon_ = epsilon[pmlthick:-pmlthick, pmlthick:-pmlthick, :]
        sigma_ = sigma[pmlthick:-pmlthick, pmlthick:-pmlthick, :]
        np.save(erdir + "/" + str(int(epoch / save_epoch)) + "sigma.npy", se_.cpu().detach().numpy())
        np.save(erdir + "/" + str(int(epoch /save_epoch)) + "epsilon.npy", er_.cpu().detach().numpy())
    # Clamp values to physical ranges to avoid unphysical results
    with torch.no_grad():
        sigma.data.clamp_(min=0.000)
        epsilon.data.clamp_(min=1)
    print("-----------")
...
\end{mycode}

\subsection{Implementation of CUDA kernel function}
The computational framework consists of six CUDA kernel files and each one is designed for a specific purpose: two for source and receiver operations, two for updating the Perfectly Matched Layer (PML) boundaries, one for electromagnetic wavefield updates, and one for parameter updates.

The forward modeling process involves the computation of six update coefficients based on material properties. For the electric wavefield update, the coefficients are defined as follows: $C_{ee}$ for electric wavefield self-update, $C_{eh}$ for the electric-magnetic coupling term, and $C_{es}$ for electric source injection. For the magnetic wavefield, the corresponding coefficients are $C_{hh}$ for magnetic wavefield self-update, $C_{he}$ for the coupling with the electric, and $C_{hs}$ for magnetic source injection. These coefficients are functions of the relative permittivity $\varepsilon_r$, electrical conductivity $\sigma$, relative permeability $\mu_r$, time step $\Delta t$, and spatial discretization $\Delta x, \Delta y, \Delta z$. The following kernel function illustrates the coefficient computation performed during forward modeling. The inversion procedure is analogous to the forward modeling, but it contains modifications.

\begin{mycode}
__global__ void ucget(
    ...
    //parameters
    ...
){                         
    // Compute the global thread index
    int idx = blockIdx.x * blockDim.x + threadIdx.x;         
    // Convert linear index to 3D index: i (x-dimension)    
    int i = idx / (NY_FIELDS * NZ_FIELDS);               
    // j (y-dimension)        
    int j = (idx % (NY_FIELDS * NZ_FIELDS)) / NZ_FIELDS; 
    // k (z-dimension)        
    int k = (idx % (NY_FIELDS * NZ_FIELDS)) % NZ_FIELDS;         
    // Ensure index is within bounds, excluding the last cell in each dimension
    if (i < (NX_FIELDS - 1) && j < (NY_FIELDS - 1) && k < (NZ_FIELDS - 1)) { 
        // Calculate coefficient HA for magnetic wavefield update based on relative permeability
        float HA = m0 * mr[INDEX3D_FIELDS(i, j, k, NY_FIELDS, NZ_FIELDS)] / dt;
        // Compute magnetic wavefield update coefficients
        uH0[INDEX3D_FIELDS(i, j, k, NY_FIELDS, NZ_FIELDS)] = 1;
        uH1[INDEX3D_FIELDS(i, j, k, NY_FIELDS, NZ_FIELDS)] = (1 / dx) * 1 / HA;
        uH4[INDEX3D_FIELDS(i, j, k, NY_FIELDS, NZ_FIELDS)] = 1 / HA;
        // If conductivity is very high (e.g. metal or PEC), skip electric wavefield updates
        if (se[INDEX3D_FIELDS(i, j, k, NY_FIELDS, NZ_FIELDS)] > 100) {
            uE0[INDEX3D_FIELDS(i, j, k, NY_FIELDS, NZ_FIELDS)] = 0;
            uE1[INDEX3D_FIELDS(i, j, k, NY_FIELDS, NZ_FIELDS)] = 0;
            uE4[INDEX3D_FIELDS(i, j, k, NY_FIELDS, NZ_FIELDS)] = 0;
        } else {
            // Compute EA and EB for electric wavefield updates based on permittivity and conductivity
            float EA = (e0 * er[INDEX3D_FIELDS(i, j, k, NY_FIELDS, NZ_FIELDS)] / dt) 
                       + 0.5 * se[INDEX3D_FIELDS(i, j, k, NY_FIELDS, NZ_FIELDS)];
            float EB = (e0 * er[INDEX3D_FIELDS(i, j, k, NY_FIELDS, NZ_FIELDS)] / dt) 
                       - 0.5 * se[INDEX3D_FIELDS(i, j, k, NY_FIELDS, NZ_FIELDS)];
            // Compute electric wavefield update coefficients
            uE0[INDEX3D_FIELDS(i, j, k, NY_FIELDS, NZ_FIELDS)] = EB / EA;
            uE1[INDEX3D_FIELDS(i, j, k, NY_FIELDS, NZ_FIELDS)] = (1 / dx) * 1 / EA;
            uE4[INDEX3D_FIELDS(i, j, k, NY_FIELDS, NZ_FIELDS)] = 1 / EA;
        }
    }
}
\end{mycode}

The following code is used to store six wavefield variables, $Ex$, $Ey$, $Ez$, $Hx$, $Hy$, $Hz$.
\begin{mycode}
__global__ void store_outputs(
    ...
    //parameters
    ...                                  
) {
    // Receiver index (parallelized across X block dimension)
    int rx = blockIdx.x * blockDim.x + threadIdx.x;                    
    // Time step index (parallelized across Y block dimension)
    int s = blockIdx.y * blockDim.y + threadIdx.y;                      
    // Ensure receiver and time step indices are within bounds
    if (rx < NRX && s < step) {
        // Retrieve receivers (i, j, k) coordinates from rxcoords for step s and receiver rx
        int i = rxcoords[INDEX3D_RXCOORDS(s, rx, 0, NRX, DIM)];
        int j = rxcoords[INDEX3D_RXCOORDS(s, rx, 1, NRX, DIM)];
        int k = rxcoords[INDEX3D_RXCOORDS(s, rx, 2, NRX, DIM)];
        // Store Ex at receiver location and time step into the output array rxs
        rxs[INDEX4D_RXS(s, 0, iteration, rx, NY_RXS, N_ITER, NRX)] = Ex[INDEX4D_FIELDS(s, i, j, k, NX, NY, NZ)];
        // Store Ey component
        rxs[INDEX4D_RXS(s, 1, iteration, rx, NY_RXS, N_ITER, NRX)] = Ey[INDEX4D_FIELDS(s, i, j, k, NX, NY, NZ)];
        // Store Ez component
        rxs[INDEX4D_RXS(s, 2, iteration, rx, NY_RXS, N_ITER, NRX)] = Ez[INDEX4D_FIELDS(s, i, j, k, NX, NY, NZ)];
        // Store Hx component
        rxs[INDEX4D_RXS(s, 3, iteration, rx, NY_RXS, N_ITER, NRX)] = Hx[INDEX4D_FIELDS(s, i, j, k, NX, NY, NZ)];
        // Store Hy component
        rxs[INDEX4D_RXS(s, 4, iteration, rx, NY_RXS, N_ITER, NRX)] = Hy[INDEX4D_FIELDS(s, i, j, k, NX, NY, NZ)];
        // Store Hz component
        rxs[INDEX4D_RXS(s, 5, iteration, rx, NY_RXS, N_ITER, NRX)] = Hz[INDEX4D_FIELDS(s, i, j, k, NX, NY, NZ)];
    }
}
\end{mycode}

The following code is the excitation of hertzian dipole source. It can be excited in three directions. The modeling code using the adjoint source is similar.
\begin{mycode}
__global__ void Update_hertzian_dipole(
    ...
    //parameters
    ...             
) {
    // Compute source index using block and thread ID
    int src = blockIdx.x * blockDim.x + threadIdx.x;              
    // Compute time step index using block and thread ID        
    int s = blockIdx.y * blockDim.y + threadIdx.y;                        
    if (src < NY_SRCINFO && s < step) {
        // Load the source position (i, j, k) from srcinfo1 array
        // X coordinate of source at time step s
        int i = srcinfo1[s * NY_SRCINFO * 3 + src * 3 + 0];              
        // Y coordinate of source at time step s 
        int j = srcinfo1[s * NY_SRCINFO * 3 + src * 3 + 1];               
        // Z coordinate of source at time step s
        int k = srcinfo1[s * NY_SRCINFO * 3 + src * 3 + 2];               
        // Scalar distance parameter for source scaling
        float dl = srcinfo2;          
        // Retrieve waveform amplitude for current source and iteration                                    
        float waveform_value = srcwaveforms[src * iterations + iteration]; 
        // Compute the scaling factor based on waveform, source size, and voxel volume
        float scale = waveform_value * dl / (dx * dy * dz);
        // Inject the scaled source into the appropriate electric wavefield component based on polarisation
        // Update Ex wavefield component
        if (polarisation == 0) {
            Ex[INDEX4D_FIELDS(s, i, j, k, NX, NY, NZ)] -= 
                uE4[INDEX4D_FIELDS(s, i, j, k, NX, NY, NZ)] * scale;      
        }
        // Update Ey wavefield component
        else if (polarisation == 1) {
            Ey[INDEX4D_FIELDS(s, i, j, k, NX, NY, NZ)] -= 
                uE4[INDEX4D_FIELDS(s, i, j, k, NX, NY, NZ)] * scale;      
        }
        // Update Ez wavefield component
        else if (polarisation == 2) {
            Ez[INDEX4D_FIELDS(s, i, j, k, NX, NY, NZ)] -= 
                uE4[INDEX4D_FIELDS(s, i, j, k, NX, NY, NZ)] * scale;      
        }
    }
}
\end{mycode}

This is the code for updating the electric wavefield, which updates the coefficients obtained from the previous ones. The code for magnetic wavefield updates is similar.
\begin{mycode}
__global__ void e_fields_updates_gpu(
    ...
    //parameters
    ...     
) {
    // Compute global thread index
    int idx = blockIdx.x * blockDim.x + threadIdx.x;
    // Total number of grid cells per time step
    int total_cells_per_step = NX_FIELDS * NY_FIELDS * NZ_FIELDS;
    // Compute time step index from global thread index
    int s = idx / total_cells_per_step;
    // Compute 3D spatial indices (i, j, k) from flattened index
    int i = idx % total_cells_per_step / (NY_FIELDS * NZ_FIELDS);
    int j = (idx % total_cells_per_step % (NY_FIELDS * NZ_FIELDS)) / NZ_FIELDS;
    int k = idx % total_cells_per_step % NZ_FIELDS;
    // Only update valid grid points (excluding i=0 and j=0 for finite difference)
    if (s < step && i > 0 && i < NX_FIELDS && j > 0 && j < NY_FIELDS && k >= 0 && k < NZ_FIELDS) {
        // Update the z-component of the electric wavefield using FDTD scheme
        Ez[INDEX4D_FIELDS(s, i, j, k, NX_FIELDS, NY_FIELDS, NZ_FIELDS)] =
            uE0[INDEX4D_FIELDS(s, i, j, k, NX_FIELDS, NY_FIELDS, NZ_FIELDS)] *  
            Ez[INDEX4D_FIELDS(s, i, j, k, NX_FIELDS, NY_FIELDS, NZ_FIELDS)] +   
            uE1[INDEX4D_FIELDS(s, i, j, k, NX_FIELDS, NY_FIELDS, NZ_FIELDS)] *  
            (Hy[INDEX4D_FIELDS(s, i, j, k, NX_FIELDS, NY_FIELDS, NZ_FIELDS)] -
             Hy[INDEX4D_FIELDS(s, i - 1, j, k, NX_FIELDS, NY_FIELDS, NZ_FIELDS)]) -
            uE1[INDEX4D_FIELDS(s, i, j, k, NX_FIELDS, NY_FIELDS, NZ_FIELDS)] * 
            (Hx[INDEX4D_FIELDS(s, i, j, k, NX_FIELDS, NY_FIELDS, NZ_FIELDS)] -
             Hx[INDEX4D_FIELDS(s, i, j - 1, k, NX_FIELDS, NY_FIELDS, NZ_FIELDS)]);
    }
}
\end{mycode}
Through the integration of these kernel functions, a high-performance computing core for ground penetrating radar (GPR) forward and adjoint modeling has been proposed using CUDA. Once the CUDA source files (*.cu) are compiled, they generate corresponding shared object (*.so) files, which can be dynamically loaded and invoked from Python.

\subsection{Python interacts with CUDA}
With the CUDA kernel functions in place, the next step involves interfacing them with Python to execute the FWI process. This is achieved by dynamically loading the compiled shared library using Python's ctypes module. The following code demonstrates how to link and call the CUDA kernels from Python to enable seamless integration and high-performance execution.
\begin{mycode}
# Construct the path to the shared library 'fields_updates_gpu.so' located in the 'lib' directory next to the current script
lib_path = os.path.join(os.path.dirname(__file__), 'lib', 'fields_updates_gpu.so')
# Load the shared library using ctypes
lib = ctypes.cdll.LoadLibrary(lib_path)
# Specify the argument types expected by the C function 'e_fields_updates' from the shared library
lib.e_fields_updates.argtypes = [
    ctypes.POINTER(ctypes.c_float),  # Placeholder for actual argument types (e.g., input/output arrays)
    ...
]
# Specify that the C function 'e_fields_updates' has no return value (void)
lib.e_fields_updates.restype = None
# Call
lib.e_fields_updates(
  ...
)  
\end{mycode}

Within each time step $\Delta t$, the forward modeling process follows the sequence outlined in Algorithm~\ref{alg:Alg1}, which is divided into seven computational stages. The adjoint (backward) modeling process, illustrated in Algorithm~\ref{alg:Alg2}, consists of six steps, as it does not require storage of the forward wavefield.
\begin{algorithm}
  \caption{Forward modeling process}
  \begin{algorithmic}
  \label{alg:Alg1}
  \State \textbf{Input:} Initialize the wavefield and update coefficients.
  \newline
   \For{ i = 1, ..., iterations}
   \State \textit{1.} Save receivers data;
   \State \textit{2.} Update magnetic wavefield;
   \State \textit{3.} Update magnetic PML;
  \State \textit{4.} Update electric wavefield;
   \State \textit{5.} Update electric PML;
   \State \textit{6.} Inject the source function;
   \State \textit{7.} Save wavefield;
   \EndFor
  \newline
\State  \textbf{Output: } Forward wavefields
  \end{algorithmic} 
\end{algorithm}

\begin{algorithm}
  \caption{Backward modeling process}
  \begin{algorithmic}
  \label{alg:Alg2}
  \State \textbf{Input:} Forward wavefield, residual, update coefficients.
  \newline
   \For{ i = iterations, ..., 1}
   \State \textit{1.} Inject the adjoint source function;
   \State \textit{2.} Update electric wavefield;
   \State \textit{3.} Update electric PML;
  \State \textit{4.} Update magnetic wavefield;
   \State \textit{5.} Update magnetic PML;
   \State \textit{6.} Calculate gradients of the relative dielectricpermittivity and electrical conductivity;
   \EndFor
  \newline
\State  \textbf{Output: } gradients of the relative dielectricpermittivity and electrical conductivity
  \end{algorithmic} 
\end{algorithm} 
\subsection{Optimization strategy}
Our optimization strategies include total variation regularization \citep{rudin1992nonlinear}, staged learning rate, gradient pruning, etc.

\textbf{Total Variation (TV) regularization} is a widely adopted technique for tasks such as denoising and image reconstruction. It is well-suited for preserving sharp edges while suppressing noise-induced oscillations \citep{strong2003edge}. In this work, TV regularization is implemented within the PyTorch framework and it can be efficiently accelerated using CUDA, ensuring minimal impact on overall runtime.
\begin{algorithm}
  \caption{Total Variation regularization}
  \begin{algorithmic}
  \label{alg:Alg3}
  \State \textbf{Input:} $epsilon$, $epsilon.grad$, $sigma$, $sigma.grad$.
  \newline
  \State \textit{1.} Initialize
   \For{ i = 1, ..., iterations}
   \State \textit{2.} Compute backward difference (circular boundary);
   \State \textit{3.} Construct RHS (Right-Hand Side);
  \State \textit{4.} Compute average of neighbors using convolution;
   \State \textit{5.} Compute forward difference (circular boundary);
   \State \textit{6.} Update Z variables via soft-thresholding;
   \State \textit{7.} Update multipliers
   \EndFor
   \State \textit{8.}Restore result to original scale
  \newline
\State  \textbf{Output: } $epsilon.grad+eposilon\_tv\_grad$, $sigma.grad+sigma\_tv\_grad$.
  \end{algorithmic} 
\end{algorithm} 
As shown in Algorithm~\ref{alg:Alg3}, the TV method introduces auxiliary variables $Z_x$ and $Z_y$, which approximate the first-order derivatives of the model in the horizontal and vertical directions, respectively. These variables are updated using a soft-thresholding function to enforce sparsity and preserve edges. The original TV regularization term is defined as:
\begin{align}
TV(X) = \sum_{i,j} \sqrt{(D_x X)^2 + (D_y X)^2}.
\end{align}
where $X$ denotes the model parameter to be reconstructed, such as relative permittivity or conductivity; $D_x$ and $D_y$ represent the discrete gradient operators in the horizontal and vertical directions, respectively. In the source code, the TV regularization can be activated by enabling the \texttt{tv} option in the configuration.

\textbf{Multiscale inversion} is an essential strategy integrated into the proposed method to improve convergence, stability, and resolution in FWI. This approach initiates the inversion process with low-frequency data and progressively incorporates higher-frequency information. By doing so, it mitigates the risk of converging to local minima and stabilizes gradient-based optimization in highly nonlinear scenarios. Moreover, multiscale inversion exploits the complementary sensitivities of different frequency bands—where low frequencies capture deep, large-scale structures and high frequencies resolve shallow, fine-scale anomalies—resulting in more accurate and hierarchically structured subsurface models. Additionally, the early-stage use of coarse grids and low-frequency data suppresses high-frequency noise and reduces computational demands, allowing the inversion to begin with a robust initial model and transition efficiently to higher resolutions with minimal overall runtime.

This functionality can be activated via the \texttt{freq} option in the parameter settings.

\textbf{Staged learning rate} scheduling is a widely adopted strategy in gradient-based optimization that involves adjusting the learning rate at predefined stages during the training or inversion process. This approach offers several advantages, particularly in the context of highly nonlinear and ill-posed problems such as FWI. Initially, a relatively large learning rate facilitates rapid convergence by enabling substantial parameter updates and effective exploration of the solution space. As the optimization progresses and approaches a local minimum, the learning rate is systematically reduced to allow finer adjustments, thereby preventing oscillations. This hierarchical adjustment improves both the convergence stability and the final inversion accuracy.

\section{Results}
All numerical experiments presented in this paper were conducted on a workstation equipped with a 12th Generation Intel® Core™ i7 processor (12 cores, 20 threads) and an NVIDIA GeForce RTX 4090 GPU (24 GB memory). The system configuration included Ubuntu 20.04 LTS, Python 3.10.14, NVIDIA driver version 535.230.02, and CUDA toolkit version 12.2.

To evaluate the effectiveness and computational efficiency of the proposed method, we conducted a series of experiments on different models. The results confirm that our method can produce reliable inversion outcomes within minutes, making it a practical tool for validating new methodologies and facilitating rapid iterative development.

\subsection{Cross-hole dual-parameter FWI}
\begin{figure}
\centering
\includegraphics[width=0.75\textwidth]{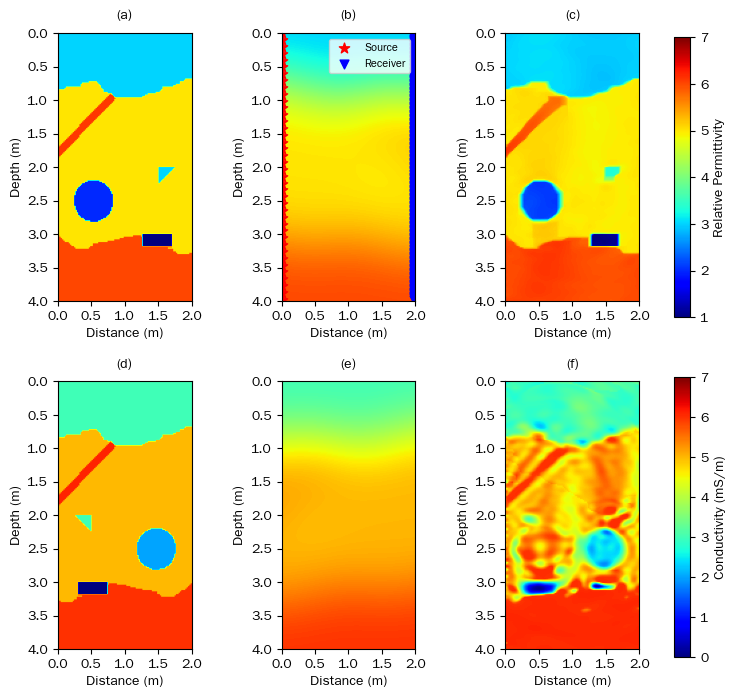}
\caption{Cross-hole dual-parameter GPR FWI. (a) true relative permittivity model; (b) initial relative permittivity model; (c) inverted relative permittivity model; (d) true conductivity model; (e) initial conductivity model; (f) inverted conductivity model.}
\label{fig:ddfwi}
\end{figure}

We constructe a landfill model with three anomalies and a leakage channel embedded in a three-layer background. The relative permittivity model and the conductivity model have different spatial distributions to test the dual-parameter inversion capability of the method. The true relative permittivity and conductivity models are shown in \cref{fig:ddfwi}(a) and \cref{fig:ddfwi}(c), respectively, while \cref{fig:ddfwi}(b) and \cref{fig:ddfwi}(e) display the corresponding initial models. Source and receiver positions are indicated in the initial model plots.

The FWI were conducted on a 2D computational domain with spatial grid spacing $\Delta x = 0.05\,\mathrm{m}$ and a temporal resolution of $\Delta t = 10^{-10}\,\mathrm{s}$. The total simulation time window was $10^{-7}\,\mathrm{s}$. The model size was set to $120 \times 220$ grid points in the horizontal and vertical directions, respectively. A single vertically oriented Hertzian dipole source was employed, with a Ricker wavelet of central frequency $f_0 = 100\,\mathrm{MHz}$. The source was positioned at $(x, y) = (0.5\,\mathrm{m}, 0.5\,\mathrm{m})$, and the receivers were deployed along a line at $y = 5.5\,\mathrm{m}$, ranging from $x = 0.5\,\mathrm{m}$ to $x = 10.45\,\mathrm{m}$ at intervals of $0.05\,\mathrm{m}$, resulting in a total of 200 receiver positions. The source was activated only once at a single location. A PML of 10 grid points thickness was applied on all domain boundaries to suppress reflections. A two-stage learning rate strategy was employed to facilitate stable and efficient convergence. In the first stage(<50 epochs), the relative permittivity $\varepsilon_r$ was optimized using the Adam optimizer with a learning rate of 0.2. In the second stage, both $\varepsilon_r$ and the electrical conductivity $\sigma$ were updated jointly, with respective learning rates set to 0.1 and 0.0001. We use this sequential inversion strategy to mitigate the corsstalk issue for dual-parameter inversion. We first update epsilon as it is the dominant parameter that controls the speed of electromagnetic waves \citep{song2020sequential}. TV regularization was applied to the model parameters to promote spatial sparsity and preserve sharp interfaces. The detailed parameter settings for this test are listed in \cref{tab:para1}.

\cref{fig:ddfwi}(c) and \cref{fig:ddfwi}(f) display the inversion results. It can be seen from the figure that the inversion result of relative permittivity is basically consistent with the real model, the outlines of several target objects are outlined, and the values are basically correct. The inversion result of conductivity is basically fine where the target objects exist, but the accuracy of the values needs to be improved. 

Our method achieves excellent computational efficiency, with a single epoch taking only 1.14 seconds and a total run time of 195.11 seconds for 150 epochs, less than 4 minutes. Despite the limited run time, the inversion results show high accuracy in reconstructing the relative permittivity distribution and clearly capture the target conductivity anomaly.

\begin{table}
\centering
\caption{Cross hole dual objective dual parameter model parameters.}
\begin{tabular}{lc}
\toprule
\textbf{Parameters} & \textbf{Value} \\
\midrule
device & cuda \\
dx  & 0.05 \\
nx & 220 \\
ny & 120 \\
dz & 1 \\
time\_windows & $1\times10^{-7}$ \\
dt & $1\times10^{-10}$ \\
source & ['hertzian', 'z'] \\
source\_amplitudes & ['ricker', 1, $1\times10^8$] \\
source\_location & [0.5, 0.50, 0] \\
receiver\_location & [0.5, 5.50, 0] \\
source\_step & [0.25, 0, 0] \\
receiver\_step & [0.05, 0, 0] \\
pmlthick & 10 \\
step & 40 \\ 
nsrc & 1 \\
nrx & 200 \\
n\_epoch & 150 \\
\bottomrule
\end{tabular}
\label{tab:para1}
\end{table}

\subsection{Surface acquisition single parameter Overthrust model FWI test}
\begin{figure}
\centering
\includegraphics[width=0.75\textwidth]{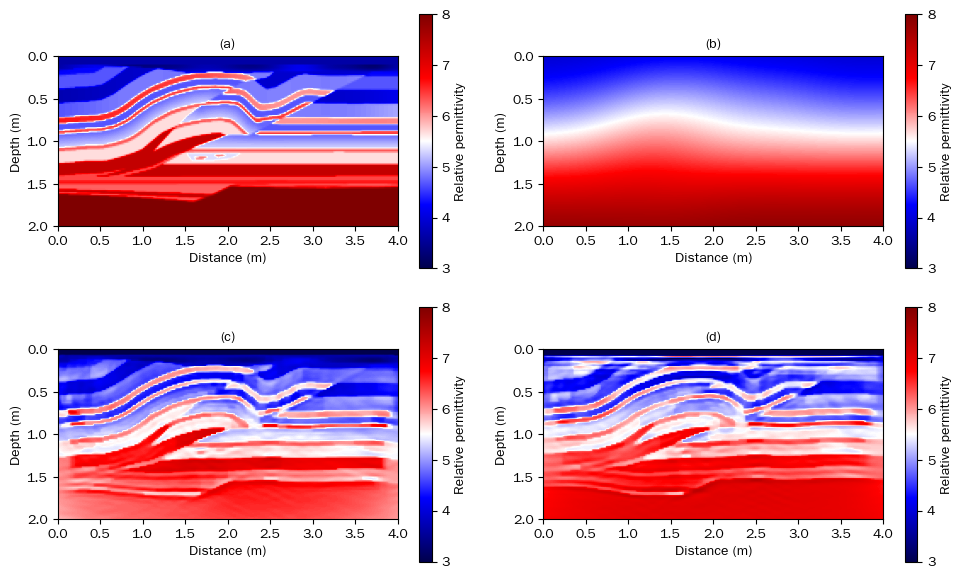}
\caption{Overthrust model. (a) true relative permittivity model; (b) initial relative permittivity model; (c) inverted relative permittivity model from the proposed method; (d) inverted relative permittivity model.}
\label{fig:otfwi}
\end{figure}

The Overthrust model, a benchmark example frequently used in seismic FWI studies, is employed to evaluate the performance of our proposed method on single-parameter inversion task. In this test, we performe FWI of relative dielectric permittivity only. The inversion results are presented in \cref{fig:otfwi}(c), which demonstrate that the model was successfully reconstructed within approximately eight minutes.

The Overthrust model inversion was conducted on a 2D computational grid with a spatial discretization of $\Delta x = 0.02\,\mathrm{m}$ and $\Delta y = 0.02\,\mathrm{m}$, comprising $220 \times 120$ grid points in the horizontal and vertical directions, respectively. The simulation used a time step of $\Delta t = 4 \times 10^{-11}\,\mathrm{s}$, and the total time window was $1.4 \times 10^{-8}\,\mathrm{s}$. A single vertical Hertzian dipole source was employed, located at $(x, y) = (0.2\,\mathrm{m}, 0.2\,\mathrm{m})$, with a Ricker wavelet of central frequency $400\,\mathrm{MHz}$ and unit amplitude. Both the source and receiver were moved along the $y$-direction in steps of $0.04\,\mathrm{m}$, although in this configuration only one source and one receiver location were used. A PML of 10 grid points was applied at each boundary to eliminate reflections. The inversion was performed 200 epochs using CUDA acceleration on GPU hardware. TV regularization was used. The overall learning rate is 0.02. The detailed parameter settings for this test are listed in \cref{tab:paraot}.

To benchmark the performance of the proposed method in terms of both accuracy and efficiency, we conducte comparative experiments using grids of the same size with an alternative CPU-based parallel FWI implementation \citep{wang2025gpr}. The inverted relative dielectric permittivity from the CPU-based method is shown in \cref{fig:otfwi}(d). The two methods run with the same inversion setup, including the grid size, number of sources, optimization algorithms, and so on. The only difference is that one runs on the CPU and the other runs on the GPU. The SSIM(Structural similarity index measure) can measure the degree of distortion in an image, as well as the degree of similarity between two images \citep{1284395}. The maximum value is 1, indicating that the two images are completely identical. The time consumption and SSIM comparisons are summarized in \cref{tab:ot}. It can be seen from \cref{tab:ot} that running a single epoch of our proposed GPU-based method is 25 times faster than the CPU-based method. 

From the comparative experiment, it can be concluded that our method has increased the speed by an order of magnitude compared to the method running on the CPU, effectively accelerating the GPR FWI.

\begin{table}
\centering
\caption{Overthrust FWI parameters.}
\begin{tabular}{lc}
\toprule
\textbf{Parameters} & \textbf{Value} \\
\midrule
device & cuda \\
dx  & 0.02 \\
nx & 120 \\
ny & 220 \\
dz & 1 \\
time\_windows & $1.4\times10^{-8}$ \\
dt & $4\times10^{-11}$\\
source & ['hertzian', 'z'] \\
source\_amplitudes & ['ricker', 1, $4\times10^8$] \\
source\_location & [0.2, 0.2, 0] \\
receiver\_location & [0.2, 0.2, 0] \\
source\_step & [0, 0.04, 0] \\
receiver\_step & [0, 0.04, 0] \\
pmlthick & 10 \\
step & 100 \\ 
nsrc & 1 \\
nrx & 1 \\
n\_epoch & 200 \\
\bottomrule
\end{tabular}
\label{tab:paraot}
\end{table}

\begin{table}
\centering
\caption{Comparison of running time.}
\label{tab:ot}
\begin{tabular}{ |c||c|c|c|c|} 
 \hline
     & Single epoch runtime ($s$)  & epoch& Total runtime ($s$) & SSIM \\ 
 \hline
 \hline
Our method & 2.17& 200 &  440.02  & 0.7277   \\
\hline
CPU-based method & 54.79   & 200& 10773.02  & 0.5512   \\
\hline
\end{tabular} 
\end{table}

\subsection{Real data dual-parameter FWI}
\begin{figure}
\centering
\includegraphics[width=0.75\textwidth]{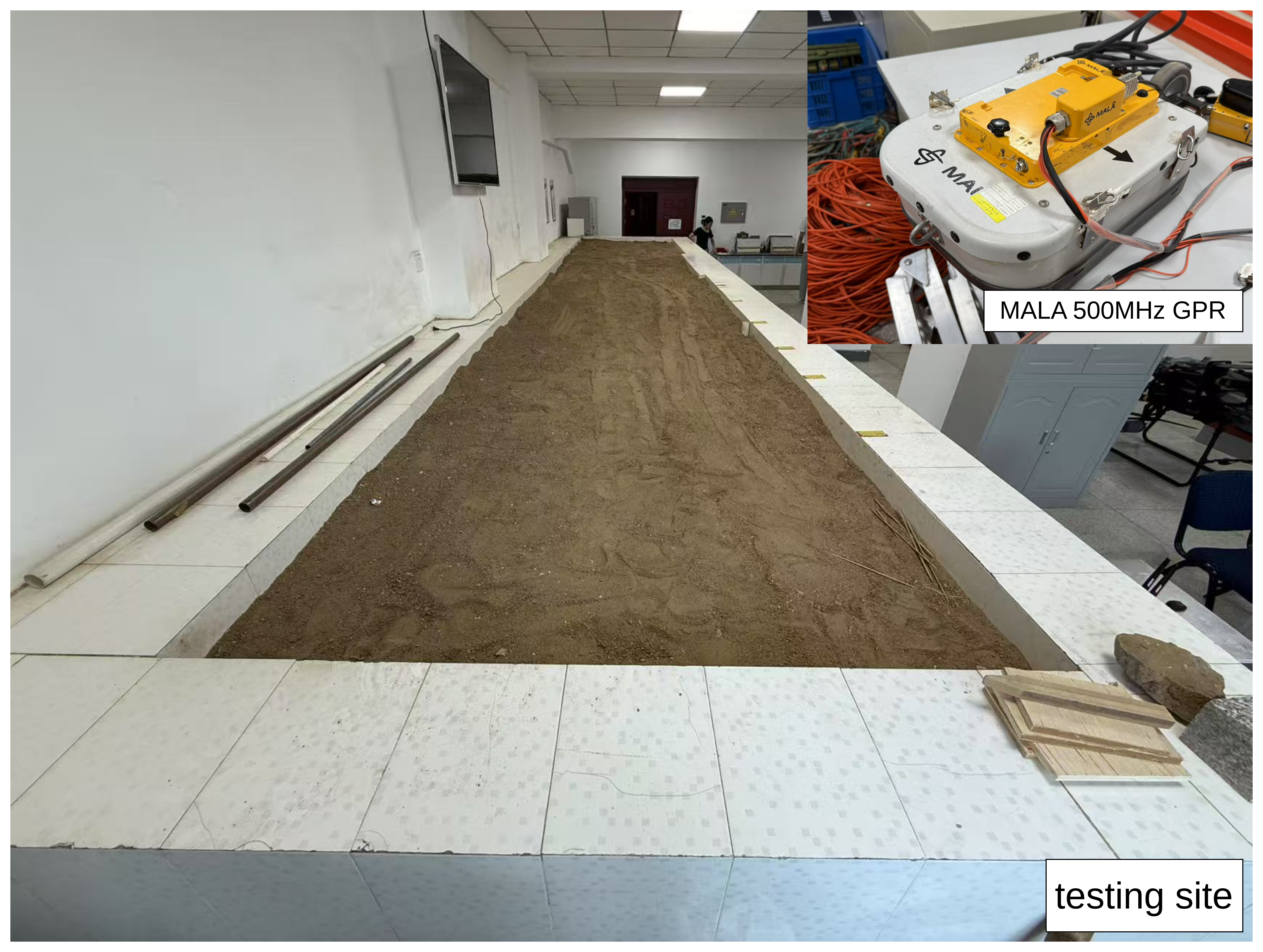}
\caption{GPR and testing site used.}
\label{fig:yq}
\end{figure}

\begin{figure}
\centering
\includegraphics[width=0.75\textwidth]{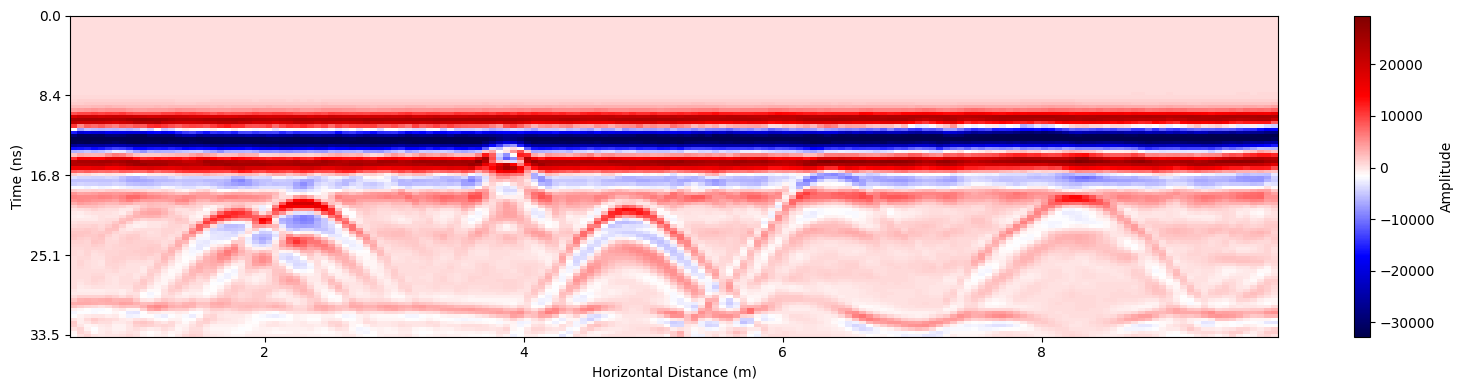}
\caption{Real observed data.}
\label{fig:realdata}
\end{figure}

This real dataset shown in \cref{fig:realdata} was acquired from a controlled pit environment filled with fine sand, containing several buried metal pipes and voids, as shown in \cref{fig:yq}. We are using a MALA 500M$Hz$ radar.

\begin{figure}
\centering
\includegraphics[width=0.75\textwidth]{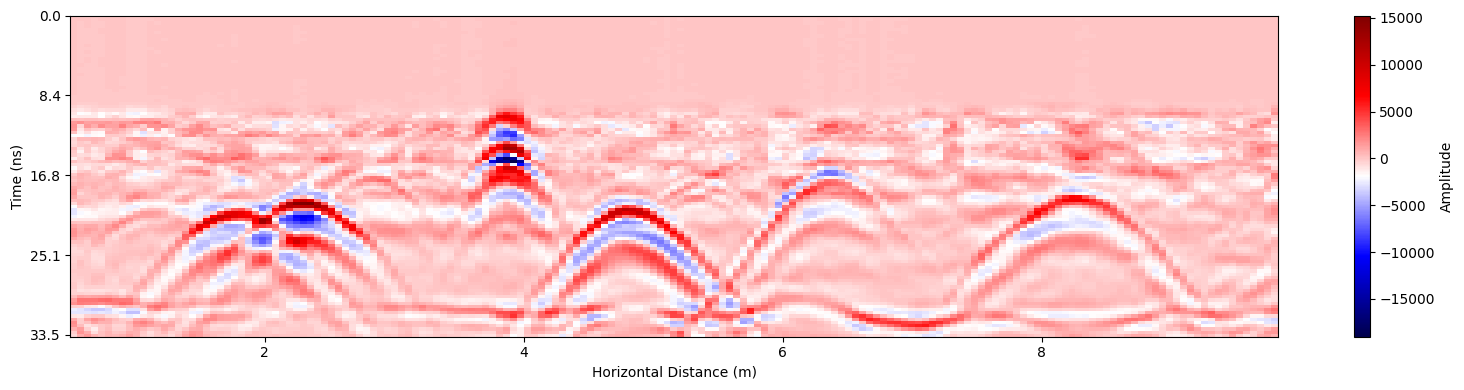}
\caption{Real observed data without direct waves.}
\label{fig:nodrealdata}
\end{figure}

For this real data test, we remove the direct wave components were removed using singular value decomposition (SVD)\citep{liu2017random} to relatively amplify the energy of the reflections, especially for signals buried in the direct waves. The processed data, with direct waves eliminated, are presented in \cref{fig:nodrealdata}.

The FWI for real observed data was carried out on a 2D domain discretized with a grid spacing of $\Delta x = 0.025\,\mathrm{m}$. The model size was set to $80 \times 374$ grid points in the horizontal and vertical directions, respectively. The total simulation time window was set to $1.943611 \times 10^{-8}\,\mathrm{s}$ with a time step of $4.871205 \times 10^{-11}\,\mathrm{s}$. A single vertically polarized Hertzian dipole source was used, emitting a Ricker wavelet with a central frequency of $500\,\mathrm{MHz}$ and unit amplitude. The source was located at $(0.25\,\mathrm{m}, 0.25\,\mathrm{m})$, and the receiver was placed at $(0.25\,\mathrm{m}, 0.425\,\mathrm{m})$, both moving in the $y$-direction with a spatial step of $0.05\,\mathrm{m}$. One source and one receiver position were actually used. To suppress boundary reflections, a PML of 10 grid points was applied on each side of the model domain. The inversion was performed 1000 epochs. All computations were executed using GPU acceleration via CUDA. TV regularization was used. The learning rate of relative permittivity is 0.1, and the learning rate of conductivity is 0.0001. The detailed parameter settings for this test are listed in \cref{tab:parar}.

The inversion was initiated from a uniform background model for both relative permittivity and electrical conductivity.

\begin{table}
\centering
\caption{Real observed data FWI parameters.}
\begin{tabular}{lc}
\toprule
\textbf{Parameters} & \textbf{Value} \\
\midrule
device & cuda \\
dx  & 0.025 \\
nx & 80 \\
ny & 374 \\
dz & 1 \\
time\_windows & $1.943611\times10^{-8}$ \\
dt & $4.871205\times10^{-11}$\\
source & ['hertzian', 'z'] \\
source\_amplitudes & ['ricker', 1, $5\times10^8$] \\
source\_location & [0.25, 0.25, 0] \\
receiver\_location & [0.25, 0.425, 0] \\
source\_step & [0, 0.05, 0] \\
receiver\_step & [0, 0.05, 0] \\
pmlthick & 10 \\
step & 173 \\ 
nsrc & 1 \\
nrx & 1 \\
n\_epoch & 1000 \\
\bottomrule
\end{tabular}
\label{tab:parar}
\end{table}

\begin{figure}
\centering
\includegraphics[width=0.75\textwidth]{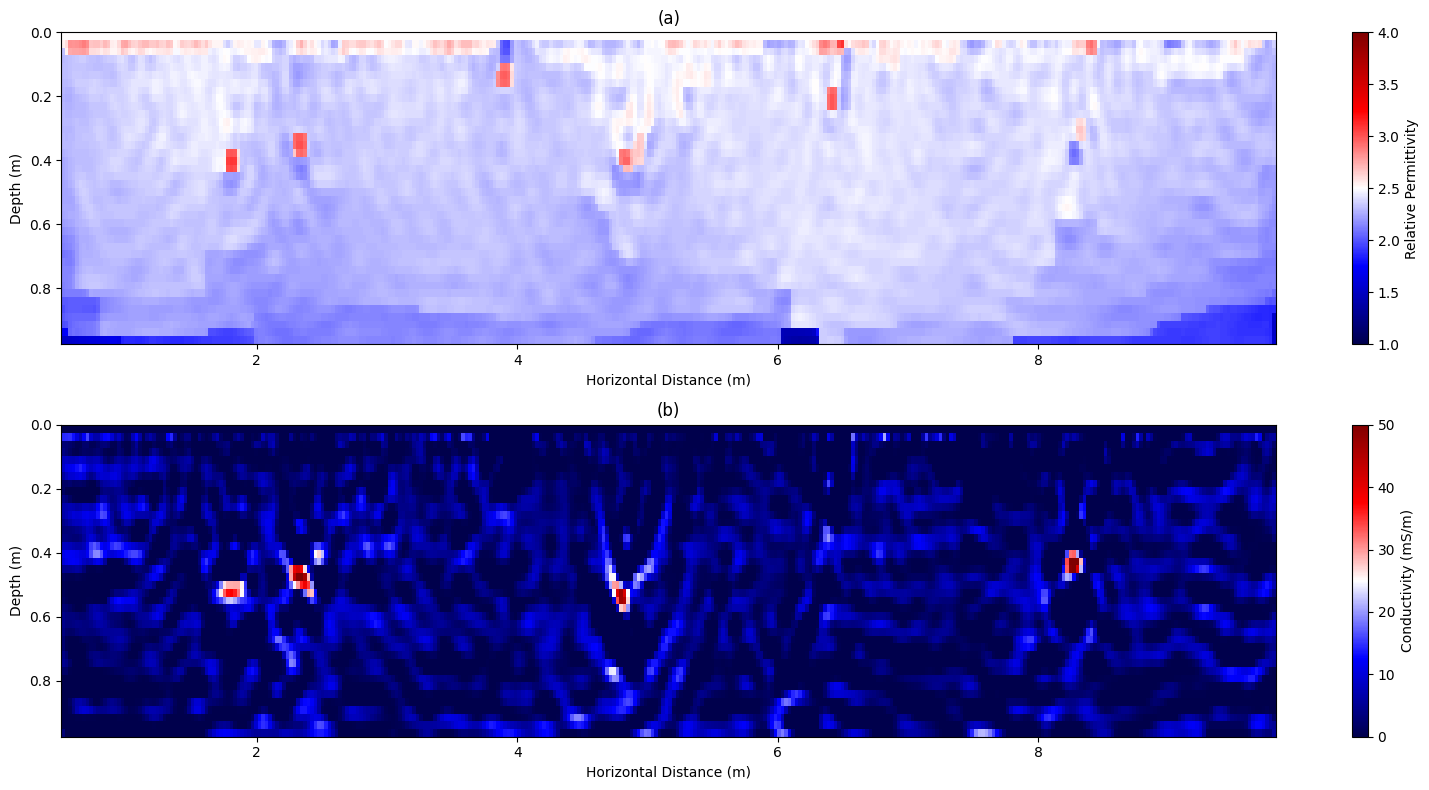}
\caption{Real data FWI results of (a) inverted relative permittivity, and (b) conductivity.}
\label{fig:realfwi}
\end{figure}

 The inversion results of relative permittivity and electrical conductivity obtained using our proposed method are displayed in Figs.\ref{fig:realfwi}(a) and \ref{fig:realfwi}(b), respectively. The results demonstrate that the proposed method achieves high accuracy in recovering subsurface features from real-world data. Specifically, six anomalous bodies are clearly captured in the relative permittivity model, while four distinct high-conductivity regions are accurately inverted in the conductivity model. These results confirm the practical effectiveness and robustness of our method in real-world GPR surveys.

\section{Conclusions and discussions}

This study proposes an accelerated full waveform inversion (FWI) framework for ground-penetrating radar (GPR), proposed through the hybrid compilation of CUDA kernel functions and PyTorch. The proposed method demonstrates high computational efficiency across both synthetic data and real datasets.

A key advantage of integrating with PyTorch lies in its support for automatic differentiation and modular design. Optimizers and loss functions from the PyTorch ecosystem can be directly utilized, eliminating the need for manual gradient derivation or custom optimization routines. Furthermore, the framework includes extensible interfaces, enabling the incorporation of additional inversion strategies and regularization techniques to further enhance reconstruction accuracy.

To facilitate future development, the codebase includes reserved interfaces for three-dimensional (3D) extension, allowing for seamless scalability from 2D to 3D modeling.

Current limitations are primarily related to GPU memory capacity, which constrains the size of the models that can be processed. Future work will address this limitation by integrating multi-GPU parallel computing and checkpointing techniques to support larger-scale inversions.

\section{Acknowledgments}

This work was supported by National Key Research and Development Program of China \\(Grant No.2023YFC3707901).

We would like to thank Dr.Richardson Alan, for constructive advice and guidance in coding. We would also like to thank Dr.Zhang Minghe from Jilin University for his help in obtaining real data.

\newpage

\textbf{Code availability section}

Fast GPR FWI

Contact: liulei24@mails.jlu.edu.cn

Hardware Requirements: Our method is designed to run on systems equipped with one or more NVIDIA GPUs. Larger GPU memory allows for the simulation and inversion of more complex or higher-resolution models. In this study, all experiments were performed on a workstation configured with a 12th Generation Intel® Core™ i7 processor (12 cores, 20 threads) and an NVIDIA GeForce RTX 4090 GPU with 24 GB of memory.

Program language: Python, C++.

Software required: PyTorch, CUDA toolkit, and some Python packages.

Program size: The compiled code is about 5.5MB (Without data).

The source codes are available for downloading at the link: https://github.com/songc0a/Fast-GPR-FWI/

\bibliographystyle{cas-model2-names}
\bibliography{bibliography} 

\end{document}